\documentstyle[multicol,aps,epsf]{revtex}

\title{Optimized Verlet-like algorithms for molecular dynamics simulations}

\author{I. P. Omelyan,$^1$ I. M. Mryglod,$^{1,2}$ and R. Folk$^2$}

\address{$^1$Institute for Condensed Matter Physics,
         1 Svientsitskii Street, UA-79011 Lviv, Ukraine}
\address{$^2$Institute for Theoretical Physics, Linz University,
         A-4040 Linz, Austria}

\date{\today}

\begin{document}

\maketitle

\begin{abstract}

New explicit velocity- and position-Verlet-like algorithms
of the second order are proposed to integrate the equations
of motion in many-body systems. The algorithms are derived
on the basis of an extended decomposition scheme at the
presence of a free parameter. The nonzero value for this
parameter is obtained by reducing the influence of truncated
terms to a minimum. As a result, the new algorithms appear
to be more efficient than the original Verlet versions which
correspond to a particular case when the introduced parameter
is equal to zero. Like the original versions, the proposed
counterparts are symplectic and time reversible, but lead to
an improved accuracy in the generated solutions at the same
overall computational costs. The advantages of the new
algorithms are demonstrated in molecular dynamics
simulations of a Lennard-Jones fluid.

\vspace{6pt}

\noindent
Pacs numbers: 02.60.Cb; 02.70.Ns; 05.10.-a; 95.75.Pq

\end{abstract}

\vspace{12pt}

\begin{multicols}{2}

The method of molecular dynamics (MD) is a powerful tool for the
prediction and study of various phenomena in physics, chemistry
and biology. In MD simulations we deal with the necessity to solve
numerically the equations of motion for a many-body system composed
of interacting particles. The most of traditional algorithms, such
as Runge-Kutta and predictor-corrector schemes \cite{Gear,Burden},
are usually unsuitable for integration of the resulting differential
equations, because the solutions obtained exhibit a high instability
on MD scales of time \cite{Allen}.

A variety of alternative algorithms were proposed and implemented
over the years \cite{Verlet,Beeman,Hockney,Swope,Frenkel}. These
include the well-known velocity-Verlet (VV) integrator \cite{Swope}.
This second-order integrator is employed in the great majority of
MD simulations due to its simplicity and exceptional stability.
Moreover, the VV algorithm is symplectic, time reversible, and
able to reach a high level of accuracy with minimal number of
force evaluations per time step \cite{Allen,Frenkel}. In addition,
the VV approach can be modified to integrate not only translational
motion in atomic systems, but also simulate more complicated molecular
and spin liquids \cite{Anders,Omcpc,Omcip,Omfes}.

The question of how to improve the efficiency of integration for
atomic systems with long-range interactions has also been
considered. As a result, so-called multiple time scale integrators
have been introduced \cite{Tuckerman,Stuart}. In these
integrators, the additional slow subdynamics is treated in a
specific way using the weakness of the long-range forces. The
faster motion, caused by the interactions at short interparticle
distances, remains to be integrated with the help of usual basic
algorithms, such as VV integrator, for instance.

In the present paper we show that even within the basic consideration
of translational motion (when additional splitting of interaction
potentials into multiple scale components is not longer allowed), the
VV algorithm presents, in fact, only a particular case among a whole
family of symplectic reversible integrators of the second-order. This
case, appears to be not so optimal and more efficient second-order
algorithms are possible.

The equations of motion for a classical system consisting of $N$
particles can be cast in the following compact form
\begin{equation}
\frac{{\rm d} {\mbox{\boldmath $\rho$}}}{{\rm d} t} =
L {\mbox{\boldmath $\rho$}}(t) \, ,
\end{equation}
where ${\mbox{\boldmath $\rho$}} \equiv \{ {\bf r}_i, {\bf v}_i \}$ denotes
the full set ($i=1,2,\ldots,N$) of phase variables with ${\bf r}_i$ and
${\bf v}_i$ being the position and velocity, respectively, of the $i$th
particle, $L$ is the Liouville operator,
\begin{equation}
L = \sum_{i=1}^N \Big( {\bf v}_i {\mbox{\boldmath $\cdot$}}
\frac{\partial}{\partial {\bf r}_i}+\frac{{\bf f}_i}{m}
{\mbox{\boldmath $\cdot$}} \frac{\partial}{\partial {\bf v}_i}
\Big) \equiv A + B \, ,
\end{equation}
${\bf f}_i=-\sum_{j (j \ne i)}^N \varphi'(r_{ij}) {\bf r}_{ij}/r_{ij}$
designates the force acting on the particles of mass $m$ each, due to
the interactions described by the potential $\varphi(r_{ij})$, and ${\bf
r}_{ij}={\bf r}_i-{\bf r}_j$. The Liouville operator has been split
in Eq.~(2) into the free-motion $A={\bf v} {\mbox{\boldmath $\cdot$}}
\partial/\partial {\bf r}$ and potential $B={\bf f}/m {\mbox{\boldmath
$\cdot$}} \partial/\partial {\bf v}$ parts with ${\bf v} \equiv \{
{\bf v}_i \}$, ${\bf r} \equiv \{ {\bf r}_i \}$, and ${\bf f} \equiv
\{ {\bf f}_i \}$.

The formal solution of Eq.~(1) is
\begin{equation}
{\mbox{\boldmath $\rho$}}(h)={\rm e}^{Lh}
{\mbox{\boldmath $\rho$}}(0) \equiv
{\rm e}^{(A+B)h} {\mbox{\boldmath $\rho$}}(0) \, ,
\end{equation}
where $h$ denotes the time step. Of course, the exponential propagator
$e^{L h}$ cannot be evaluated exactly at any $h$ (solutions in quadratures
are possible only for $N=2$ that is not relevant for our consideration of
many-body systems when $N \gg 1$). However, at small enough values of $h$,
the total propagator allows to be decomposed \cite{Yoshida,Forest,Suzukium,%
Suzuki} as
\begin{equation}
{\rm e}^{(A+B)h} = \prod_{p=1}^{P} {\rm e}^{A a_p h} {\rm e}^{B b_p h}
+ {\cal O}\big(h^{K+1}\big) \, ,
\end{equation}
where the coefficients $a_p$ and $b_p$ are chosen in such a way to provide
the highest possible value for $K \ge 1$ at a given integer number $P \ge
1$. Then, starting from an initial configuration ${\mbox{\boldmath $\rho$}}
(0)$, the evolution of the system can be investigated during arbitrary
times $t$ by repeating the single-step propagation, ${\mbox{\boldmath
$\rho$}}(t) = \big( {\rm e}^{Lh} \big)^l {\mbox{\boldmath $\rho$}}(0)
\equiv \big( {\rm e}^{(A+B) h} \big)^l {\mbox{\boldmath $\rho$}}(0)$, i.e.,
\begin{equation}
{\mbox{\boldmath $\rho$}}(t) =
\Bigg( \prod_{p=1}^{P} {\rm e}^{A a_p h}
{\rm e}^{B b_p h} \Bigg)^l {\mbox{\boldmath $\rho$}}(0) \equiv S(t)
{\mbox{\boldmath $\rho$}}(0) \, ,
\end{equation}
where $l=t/h$ is the total number of steps and the truncation terms
${\cal O}(h^{K+1})$, appearing in Eq.~(4), have been neglected.

The main advantage of decomposition (4) is that the exponential
subpropagators ${\rm e}^{A \tau}$ and ${\rm e}^{B \tau}$ are analytically
integrable. Indeed,
\begin{eqnarray}
&&{\rm e}^{A \tau} {\mbox{\boldmath $\rho$}} \equiv
{\rm e}^{{\bf v} {\mbox{\boldmath $\cdot$}} \partial/\partial {\bf r}
\tau} \{ {\bf r}, {\bf v} \} = \{ {\bf r} + \tau {\bf v}, {\bf v} \} \, ,
\nonumber \\ [-5pt] \\ [-5pt]
&&{\rm e}^{B \tau} {\mbox{\boldmath $\rho$}} \equiv {\rm e}^{{\bf
f}/m {\mbox{\boldmath $\cdot$}} \partial/\partial {\bf v} \tau} \{
{\bf r}, {\bf v} \} = \{ {\bf r}, {\bf v} + \tau {\bf f}/m \} \nonumber
\end{eqnarray}
that represents simple shifts of positions and velocities, respectively
(with $\tau$ being equal to $a_p h$ or $b_p h$). In addition, the generated
trajectory (5) behaves symplectically (like exact solutions), because
such separate shifts do not change the volume in phase space. The time
reversibility $S(-t) {\mbox{\boldmath $\rho$}}(t) = {\mbox{\boldmath
$\rho$}}(0)$ of solutions (following from the property $S^{-1}(t)=S(-t)$
of evolution operators) can also be obtained by imposing additional
conditions on the coefficients $a_p$ and $b_p$. Namely, $a_1=0$,
$a_{p+1}=a_{P-p+1}$, and $b_p=b_{P-p+1}$, or $a_p=a_{P-p+1}$, and
$b_p=a_{P-p}$ at $b_P=0$. In other words, the subpropagators ${\rm e}^{A
\tau}$ and ${\rm e}^{B \tau}$ should enter symmetrically in the
decompositions. Then the even-order truncation terms ${\cal O}(h^{2k})$
will disappear automatically in Eq.~(4) at $k \le K/2$. For this reason,
the order $K$ of reversible algorithms may accept only even numbers. The
cancellation of odd-order terms ${\cal O}(h^{2k-1})$ will be provided by
satisfying a set of basic conditions for $a_p$ and $b_p$. For example,
the condition $\sum_{p=1}^P a_p=\sum_{p=1}^P b_p=1$ is required to cancel
the first-order truncation uncertainties.

The method just highlighted is quite general to build numerical integrators
of arbitrary orders. In particular, the second-order ($K=2$) VV algorithm
\begin{equation}
{\rm e}^{(A+B) h} = {\rm e}^{B \frac{h}{2}} {\rm e}^{A h}
{\rm e}^{B \frac{h}{2}} + {\cal O}(h^3)
\end{equation}
is immediately reproduced from Eq.~(4) at $P=2$ and $a_1=0$, $b_1=b_2=1/2$,
$a_2=1$. The case when the operators $A$ and $B$ are replaced by each other
($A \leftrightarrow B$) is also possible, and we come to the so-called
position-Verlet (PV) algorithm \cite{Tuckerman}, ${\rm e}^{(A+B)h} = {\rm
e}^{A h/2} {\rm e}^{B h} {\rm e}^{A h/2} + {\cal O}(h^3)$, corresponding
to the choice $a_1=a_2=1/2$, $b_1=1$, and $b_2=0$. Algorithms of higher
orders can also be derived in a similar way. For instance, the fourth-order
($K=4$) algorithm by Forest and Ruth \cite{Forest} is obtained from Eq.~(4)
at $P=4$, whereas sixth-order ($K=6$) schemes are derivable \cite{Yoshida}
beginning from $P=8$. The high-order schemes involve, however, too large
number of force recalculations, and appear to be less efficient in most
of MD applications than second-order algorithms.

Despite the fact that the method of construction of
time-reversible integrators using symplectic decompositions is not
new, some important cases have never been considered and have been
completely ignored in the literature. This concerns, in
particular, the following question. Are the above Verlet
algorithms optimal in view of the time efficiency among all
possible basic (i.e. with single splitting of the Liouville
operator) decomposition integrators of the second-order? We can
say only that the Verlet algorithms do minimize the number of
force evaluations per time step. However, as will be shown below,
this does not guarantee the optimization with respect to the
overall number of force recalculations (the most time-consuming
part of MD simulations), which are necessary to perform during a
fixed observation time in order to achieve a given precision in
solutions.

It can be seen readily that the Verlet algorithms ($P=2$) require only
one ($P-1$) force evaluations per time step $h$, whereas the fourth- and
higher-order schemes $P \ge 4$ need in three or more such evaluations.
Let us consider now the intermediate case $P=3$ which leads to an
extended time-reversible propagation in the form
\begin{equation}
{\rm e}^{(A+B) h} = {\rm e}^{A \xi h} {\rm e}^{B \frac{h}{2}}
{\rm e}^{A (1-2\xi) h} {\rm e}^{B \frac{h}{2}} {\rm e}^{A \xi h}
+ C h^3 + {\cal O}(h^4)
\end{equation}
following from Eq.~(4) at $a_1=a_3 \equiv \xi$, $a_2=1-2\xi$, $b_1=b_2=
1/2$, and $b_3=0$. Again, the propagation with $A \leftrightarrow B$ is
also acceptable (then $a_1=0$, $b_1=b_3 \equiv \xi$, $b_2=1-2\xi$ and
$a_2=a_3=1/2$). Formula (8) represents a whole family of symplectic
time-reversible integrators of the second-order in which a particular
member can be extracted by choosing a value for free parameter $\xi$.
For $\xi=0$, Eq. (8) reduces to the VV (see Eq.~(7)) or PV (at $A
\leftrightarrow B$) algorithm. The extended (when $\xi \ne 0$) propagation
requires already two, instead of one, force recalculation per time step.
For this reason, we can come to an incorrect conclusion that such a
propagation has no advantage over the Verlet algorithms.

In order to prove that the above conclusion is indeed incorrect, let us
analyze in more detail the influence of truncation errors $C h^3$ on the
result. Expanding both the sides of Eq.~(8) into Taylor's series with
respect to $h$, one finds
\begin{equation}
C = \alpha(\xi) [A,[B,A]] + \beta(\xi) [B,[B,A]] \, ,
\end{equation}
where
\begin{equation}
\alpha(\xi) = \frac{1-6\xi+6\xi^2}{12}
\, , \ \ \ \ \ \ \
\beta(\xi) = \frac{1-6\xi}{24}
\end{equation}
and $[ \ , \ ]$ denotes the commutator of two operators. The norm of $C$
with respect to the third-order commutators $[A,[B,A]]$ and $[B,[B,A]]$
is
\begin{equation}
\gamma(\xi)=\sqrt{\alpha^2(\xi)+\beta^2(\xi)} \, .
\end{equation}
Then the norm of local uncertainties $C {\mbox{\boldmath $\rho$}} h^3$
appearing in phase trajectory ${\mbox {\boldmath $\rho$}}$ during a
single-step propagation given by Eqs.~(3) and (8) can be expressed in
terms of $\gamma$ and $h$ as $g=\gamma h^3$. During a whole integration
over a fixed time interval $t$, the total number $l$ of such single steps
is proportional to $h^{-1}$ (see Eq.~(5)). As a result, the local
third-order uncertainties will accumulate step by step leading at $t
\gg h$ to the second-order global errors $\Gamma=g h^{-1}$, i.e.,
\begin{equation}
\Gamma(\xi,h) = \gamma(\xi) h^2 \, .
\end{equation}

Extended propagation (8) can now be optimized with respect to $\xi$ by
minimizing the function $\gamma(\xi)$. As can be verified readily, the
minimum of $\gamma(\xi)$ is achieved at $\xi=\zeta$, where
\begin{eqnarray}
\zeta &=&
  \frac12 -
  \frac{\big(2 \sqrt{326}+36\big)^{1/3}}{12}
+ \frac{1}{6 \big(2 \sqrt{326}+36\big)^{1/3}}
\nonumber
\\ [4pt]
&\approx& 0.1931833275037836
\end{eqnarray}
and consists $\gamma(\zeta) \approx 0.00855$. On the other hand, the
value $\gamma(0)$ of $\gamma$ corresponding to the Verlet algorithms
(when $\xi=0$) is equal to $\gamma(0) \approx$ 0.0932, i.e., it increases
in $\gamma(0)/\gamma(\zeta) \approx 11$ times. Remembering that the
extended propagation requires two force evaluation per time step $h$, it
should be performed with double step size $2h$ with respect to that of
the Verlet algorithms, in order to provide the same number of total
force recalculations during the integration over the same time interval.
Therefore, the extended propagation will be more efficient if the following
inequality $\Gamma(\xi,2h) < \Gamma(\xi=0,h)$ takes place. Taking into
account Eq.~(12), such an inequality can be rewritten as $\gamma(0)/
\gamma(\xi) > 4$, and thus it is fulfilled completely in the optimization
regime. In particular,
\begin{equation}
\frac{\Gamma(\zeta,2h)}{\Gamma(\xi=0,h)} \approx 0.367 \, ,
\end{equation}
indicating that the optimized propagation, being applied even with double
sizes of the time step, will reduce the global errors approximately in
three times.

In view of Eqs.~(3), (6), and (8), more explicit expressions for the
single-step propagation of position and velocity from time $t$ to $t+h$
within the optimized VV-like algorithm are:
\begin{eqnarray}
{\bf r}_1&=&{\bf r}(t)+{\bf v}(t) \xi h
\nonumber \\ [1pt]
{\bf v}_1&=&{\bf v}(t)+{\textstyle\frac1m}{\bf f}[{\bf r}_1] h/2
\nonumber \\ [1pt]
{\bf r}_2&=&{\bf r}_1+{\bf v}_1 (1-2\xi) h
\\ [1pt]
{\bf v}(t+h)&=&{\bf v}_1+{\textstyle\frac1m}{\bf f}[{\bf r}_2] h/2
\nonumber \\ [1pt]
{\bf r}(t+h)&=&{\bf r}_2+{\bf v}(t+h) \xi h \, ,
\nonumber
\end{eqnarray}
whereas the optimized PV-like algorithm (when $A \leftrightarrow B$
in Eq.~(8)) reads:
\begin{eqnarray}
{\bf v}_1&=&{\bf v}(t)+{\textstyle\frac1m}{\bf f}[{\bf r}(t)] \xi h
\nonumber \\ [1pt]
{\bf r}_1&=&{\bf r}(t)+{\bf v}_1 h/2
\nonumber \\ [1pt]
{\bf v}_2&=&{\bf v}_1+{\textstyle\frac1m}{\bf f}[{\bf r}_1] (1-2\xi) h
\\ [1pt]
{\bf r}(t+h)&=&{\bf r}_1+{\bf v}_2 h/2
\nonumber \\ [1pt]
{\bf v}(t+h)&=&{\bf v}_2+{\textstyle\frac1m}{\bf f}[{\bf r}(t+h)] \xi h
\, , \nonumber
\end{eqnarray}
where the parameter $\xi$ should take its optimal value $\zeta$ (see
Eq.~(13)). The algorithms are simple, require only slight modification
with respect to the original Verlet versions, and can easily be
implemented in program codes.

It is worth pointing out that the order of local errors ${\cal O}(h^3)
\equiv C {\mbox{\boldmath $\rho$}} h^3 \equiv C \{ {\bf r}, {\bf v} \} h^3$
remains three in both position ${\bf r}(t+h)$ and velocity ${\bf r}(t+h)$
for both the optimized algorithms (15) and (16) (because the functions
$\alpha(\xi)$ and $\beta(\xi)$ cannot tend to zero simultaneously at any
$\xi$). Note also that the minimization of third-order uncertainties $C
h^3$ in Eq.~(8) automatically minimizes the fourth-order truncation terms
${\cal O}(h^4)$ which are connected with $C$ by the relation ${\cal O}
(h^4)= \frac12 \big( C (A+B) + (A+B) C \big) h^4 +{\cal O}(h^5)$. Further
optimization is also possible in specific cases. For instance, some MD
applications are aimed exclusively at the investigation of structural
properties of the system. Then the accuracy of determining particle
positions will play more important role than that of velocities. In such
a situation, it is quite natural to increase the precision in evaluation
of ${\bf r}(t+h)$ by reducing the position part $C {\bf r} h^3=\big(
\alpha(\xi) C_1 + \beta(\xi) C_2 \big) {\bf r} h^3$ of third-order
truncation errors to zero, where $C_1=[A,[B,A]]$ and $C_2=[B,[B,A]]$
(see Eq.~(9)). This reduction can be realized, because (as can be shown
using the explicit expressions for $A$ and $B$) the operator $C_2$
vanishes when acting on position, i.e. $C_2 {\bf r} = 0$, whereas $C_1
{\bf r} \ne 0$ (as well as $C_1 {\bf v} \ne 0$ and $C_2 {\bf v} \ne 0$).
The influence of $C_1 {\bf r}$ can be reduced to zero also by choosing
such $\xi$ at which $\alpha(\xi)=0$. Among the two roots $(1 \mp 1/
\sqrt{3})/2$ of equation $\alpha(\xi)=0$, the preference should be given
to the first of them, $(1-1/\sqrt{3})/2$, because it leads to a smaller
value for $\beta(\xi)$. Then replacing $\xi$ by $(1-1/\sqrt{3})/2$ in
Eq.~(15), we come to a positionally optimized VV-like algorithm in which
the positions will be generated up to the fourth-order truncation
uncertainties ${\cal O}(h^4)$.

Another useful application of the positionally optimized algorithm is
the case of weakly interacting systems, where the Liouville operator
can be presented in the form $L = A + \epsilon B$ with $\epsilon \ll 1$.
Then the operator $[B,[B,A]] \equiv C_2$, which forms the third-order
errors in velocity, will be proportional to $\epsilon^2$ and, thus,
can be neglected. For the same reason, the corresponding fourth-order
uncertainties $\frac12 \big( C_2 (A+B) + (A+B) C_2) \big) h^4$ will also
behave like $\epsilon^2$. In such a case, the positionally optimized
algorithm can be considered as a quasi fourth-order integrator which,
contrary to the usual fourth-order schemes, will require only two
(instead of three) force evaluations per time step.

In order to obtain a positionally optimized algorithm within the PV-like
integration (16), it is necessary simply to replace $\xi$ by the root
$1/6$ of equation $\beta(\xi)=0$. Note that the values $1/6 \approx 0.167$
and $(1-1/\sqrt{3})/2 \approx 0.211$ are close enough to the optimal
solution (13) which minimizes the total position-velocity uncertainties.
Nevertheless, the positionally optimized algorithms are not recommended
to be used in general case when both the position and velocity must be
evaluated with a maximal accuracy. In other words, in such partially
optimized algorithms the increased precision in position evaluation
is achieved at the expense of decreasing accuracy in determining the
velocities. Indeed, $\gamma(0)/\gamma(1/6) \equiv \gamma(0)/|\alpha(1/6)|
\approx 7$ and $\gamma(0)/\gamma((1-1/\sqrt{3})/2) \equiv \gamma(0)/
|\beta((1-1/\sqrt{3})/2)| \approx 8$ that is less than the factor
$\gamma(0)/\gamma(\zeta) \approx 11$ corresponding to optimal value (13).

Our theoretical predictions were verified by testing the VV- and PV-like
algorithms in MD simulations of a Lennard-Jones fluid (LJ). We considered
a system composed of $N=256$ particles interacting through the LJ potential
$\Phi(r)= 4 u \big[(\sigma/r)^{12}-(\sigma/r)^6\big]$ in a basic cubic box
of volume $V=L^3$ using periodic boundary conditions. The LJ potential was
truncated at $r_{\rm c}=L/2 \approx 3.36 \sigma$ and shifted to be zero at
the truncation point to avoid the force singularities, i.e., $\varphi(r)=
\Phi(r)-\Phi(r_{\rm c})$ at $r < r_{\rm c}$ and $\varphi(r)=0$ otherwise.
The simulations were carried out in a microcanonical ensemble at a reduced
density of $n^\ast=\frac{N}{V} \sigma^3=0.845$ and a reduced temperature
of $T^\ast=k_{\rm B}T/u=1.7$. The equations of motion were integrated with
the help of Eqs.~(15) and (16) in which the parameter $\xi$, being constant
within each run, varied from one run to another. All the runs started from
an identical well equilibrated initial configuration ${\mbox{\boldmath
$\rho$}}(0)$, and further continued $l=10\,000$ time steps. The precision
of generated solutions was measured in terms of the relative total energy
fluctuations ${\cal E}=\langle (E- \langle E \rangle)^2 \rangle/|\langle E
\rangle|$, where $E=\frac12 \sum_{i=1}^N m {{\bf v}_i}^2 + \frac12
\sum_{i \ne j}^N \varphi(r_{ij})$ and $\langle \ \rangle$ denotes the
microcanonical averaging. Note that in microcanonical ensembles the total
energy is an integral of motion, $E(t) = E(0)$, and the above fluctuations
should be equal to zero if the equations of motion are solved exactly. So
that in approximate MD integration, smaller values of ${\cal E}$ will
correspond to a more precise evaluation of phase trajectory.

The total energy fluctuations obtained in the simulations at the end
of the runs for four (fixed within each run) undimensional time steps,
$h^\ast=h (u/m\sigma^2)^{1/2}=0.01$, 0.005, 0.0025, and 0.001, are shown
in Fig.~1 as depending on free parameter $\xi$. The subsets (a) and (b)
of this figure correspond to the VV- and PV-like integration, respectively.
As can be seen, all the dependencies ${\cal E}(\xi,h)$ have one minimum
which locates at the same point $\xi \approx 0.19$ independently on the
size $h$ of the time step. This point coincides completely with the minimum
at $\zeta$ (Eq.~(13)) of function $\gamma(\xi)$ (Eq.~(11)) which is included
in Fig.~1 as well (dashed curves in the subsets). Moreover, the energy
fluctuations ${\cal E}(\xi,h)$ appear to be proportional to the norm
$\Gamma(\xi,h)$ of global errors (see Eq.~(12)), and the coefficient of
this proportionality

\begin{figure}[htbp]
\begin{centering}
\begin{picture}(84,103)
\epsfxsize=84mm
\put(0,0){
\epsffile[41 500 557 702]{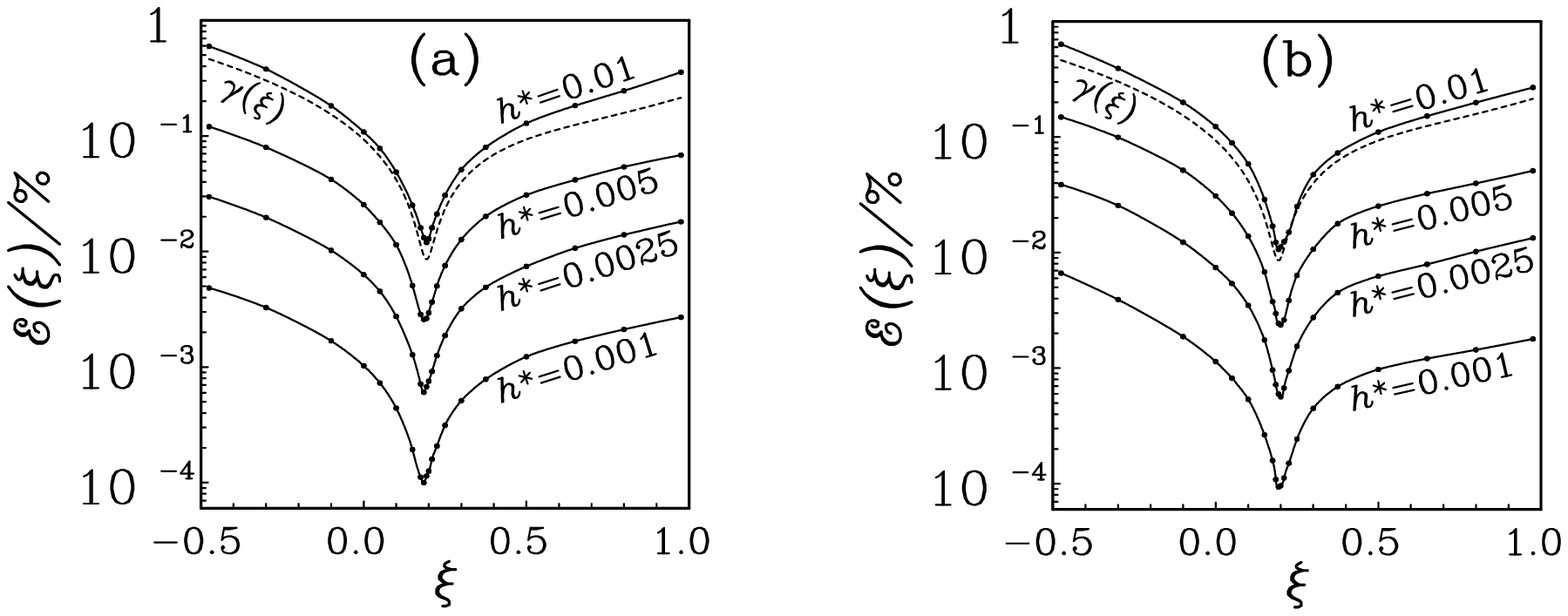}}
\end{picture}
\end{centering}
\end{figure}
{\small
FIG. 1. The total energy fluctuations obtained in the simulations for
different values of free parameter $\xi$ at four reduced time steps,
$h^\ast=0.01$, 0.005, 0.0025, and 0.001, using the VV- (subset (a)) and
PV- (subset (b)) like integration (Eqs.~(15) and (16), respectively).
The simulation results are presented by circles connected by the solid
curves. The function $\gamma(\xi)$ (see Eq.~(11)) is plotted in both
the subsets by the dashed curve.}

\vspace{12pt}

\noindent
almost does not depend on $\xi$ and $h$. In
addition, at each step size considered the energy fluctuations decrease
at the minimum more than in ten times with respect to those at $\xi=0$,
that is in agreement with our predicted value $\gamma(0)/\gamma(\zeta)
\approx 11$.

The result for the total energy fluctuations as functions of the length of
the simulations corresponding to the optimized (at $\xi = \zeta$) VV- and
PV-like algorithms is presented in Fig.~2 at the same set of time steps.
These functions are plotted by curves marked as OVV and OPV, respectively.
For the purpose of comparison, the functions corresponding to the original
VV and PV integrators are also drawn there (curves marked as VV and PV).
Note that for the original integrators, the time step within each subset
was chosen to be always twice smaller than that of the optimized versions
(this condition is necessary to provide the same number of force
recalculations during the same observation time), namely, $h^\ast=0.005$,
0.0025, 0.00125, and 0.0005 (see subsets (a), (b), (c), and (d),
respectively). Note also that within the original Verlet algorithms, the
third- and higher-orders truncation uncertainties become too big at step
sizes $h^\ast > 0.005$. In particular, then the ratio of the total energy
fluctuations to the fluctuations in potential energy (the standard ratio
for estimating the precision of the calculations) appears to be more than
a few per cent. For this reason, such large step sizes cannot be used in
precise MD simulations and, thus, are not considered in the present study.

As we see from Fig.~2, both the original (VV and PV) and optimized (OVV
and OPV) algorithms exhibit very good stability properties (the excellent
stability can be explained \cite{Allen,Frenkel} by the symplecticity and
time reversibility of the produced solutions). No systematic deviations
in the total energy fluctuations can be observed for all the integrators.
Instead, in each of the cases the amplitude of these deviations tends to
its own value which does not increase with further increasing the length
of the simulations. However, this value appears to be significantly larger
for the original versions VV and PV. On the other hand, using the optimized
OVV and OPV algorithms even with double sizes of the time step allows us to
decrease the unphysical energy fluctuations approximately in factor three.
This is in an excellent accord with our theoretical prediction (14). Note
also that the OPV algorithm is slightly better in energy conservation than
its OVV version (whereas the VV integrator is better with respect to the
PV counterpart). Furthermore, in view of the structure Eqs.~(15) and (16),
the OPV algorithm is more convenient when averaging macroscopic quantities.
In particular, then the interparticle potentials can be calculated at the
end of time steps simultaneously with the interparticle forces within the
same loop, increasing the time efficiency of the computations.

\begin{figure}[htbp]
\begin{centering}
\begin{picture}(84,195)
\epsfxsize=84mm
\put(0,0){
\epsffile[27 283 564 702]{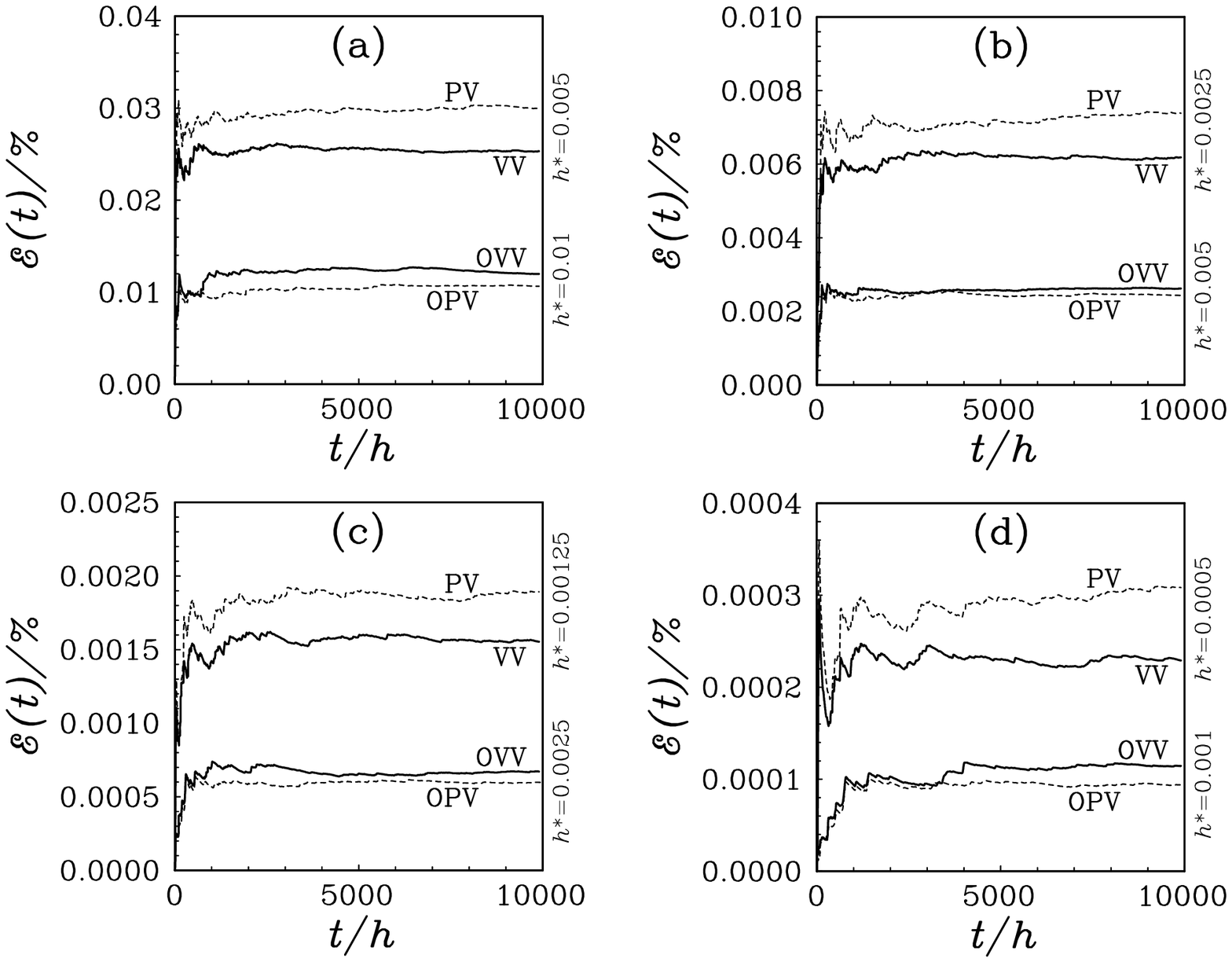}}
\end{picture}
\end{centering}
\end{figure}
{\small
FIG. 2. The total
energy fluctuations as functions of the length of the
simulations performed using the optimized VV- (solid curve marked as OVV)
and PV- (dashed curve, OPV) algorithms, as well as the original VV (solid
curve, VV) and PV (dashed curve, PV) integrators. The results corresponding
to different values of the time step, namely, $h^\ast=0.01$ and 0.005,
0.005 and 0.0025, 0.0025 and 0.00125, as well as 0.001 and 0.0005 are
presented in subsets (a), (b), (c), and (d), respectively.}

\vspace{12pt}

In conclusion, we point out that new second-order velocity- and
position-Verlet-like algorithms have been proposed to improve the
efficiency in MD simulations of classical systems. The algorithms
are explicit (i.e., do not require any iteration), simple in
implementation, and produce stable solutions which (like exact phase
trajectories) are symplectic and time reversible. The main advantage
of the introduced algorithms with respect to the widely used Verlet
integrators is the possibility to generate more precise trajectories
at the same overall computational efforts. As has been demonstrated
in a particular case of microcanonical MD simulations of a LJ fluid,
the new algorithms allow one to reduce in several times the unphysical
fluctuations of the total energy.

It has been shown rigorously within a consequent theoretical approach
that the proposed algorithms with respect to their time efficiency
should be considered as optimal among all decomposition second-order
integrators at single splitting of the Liouville operator. The optimized
algorithms can be adapted to multiple scale integration (at the presence
of long-range interactions when the potential part of the Liouville
operator is decomposed additionally) and extended to many-component
systems with orientational degrees of freedom. Moreover, the presented
decomposition (8) of noncommutative operators is applicable for quantum
Monte-Carlo simulations \cite{Suzuki} (because all the time coefficients
at the exponential propagators remain positive in the optimized regime).
These and other questions will be considered in further investigations.

Part of this work was supported by the Fonds zur F\"orderung der
wissenschaftlichen Forschung under Project P15247-TPH. I.M. and
I.O. thank the Fundamental Researches State Fund of the Ministry
of Education and Science of Ukraine for support under project
No~02.07/00303.

\end{multicols}
\end{document}